\documentclass[review]{elsarticle}

\usepackage{lineno,hyperref}
\usepackage{amsmath}
\modulolinenumbers[5]

\journal{Experimental Thermal and Fluid Science}

\begin{document}

\begin{frontmatter}

\title{Experimental and numerical analysis of grid generated turbulence with and without mean strain}

\author{J.P. Panda \fnref{myfootnote}} \author{A. Mitra} \author{A. P. Joshi} \author{H.V. Warrior}
\address{Department of Ocean Engineering and Naval Architecture,
IIT Kharagpur, India}
\fntext[myfootnote]{corresponding author: jppanda@iitkgp.ac.in}






\begin{abstract}
This paper presents experimental and numerical analysis of grid
generated turbulence with and without the effects of applied mean strain. We conduct a series of experiments on decaying grid generated turbulence and grid turbulence with mean strain. Experimental data of turbulence statistics including Reynolds stress anisotropies is collected, analyzed and then compared to the predictions of Reynolds Stress Models to assess their accuracy. The experimental data is used to evaluate the variability in the coefficients of the rate of dissipation model and the pressure strain correlation models used in Reynolds Stress Modeling. For both models we recommend optimal values of coefficients that should be used for experimental studies of grid generated turbulence.
\end{abstract}


\end{frontmatter}


\section{Introduction}
Turbulent flows appear in problems of interest to many fields of engineering sciences such as aeronautics, mechanical, chemical engineering and in oceanographic, meteorological and astrophysical sciences, besides others. Improved understanding of turbulence evolution would lead to important advances in these fields. 

At present there are no analytical solutions to predict the evolution of complex engineering turbulent flows. Studies of turbulence have to use turbulence models that characterize the statistical evolution of turbulence. Industrial studies use simple eddy viscosity based turbulence models like the $k-\epsilon$ and $k-\omega$  models. Recent emphasis in the scientific research community has shifted to Reynolds stress models \cite{gerolymos2010, klifi2013, mishra3, jakirlic2015, manceau2015, moosaie2016, sun2017,bois2017}. Reynolds stress models have the potential to give better predictions than turbulent viscosity based models at a reasonable computational expense. They may be able to model the directional effects of Reynolds stresses and complex interactions in turbulent flows \cite{mishra2,hanjalic2011}. They have the potential to accurately model the return to isotropy of decaying turbulence and evolution in the rapid distortion limit \cite{mishra6, durbin2017}. Reynolds stress models are used to develop improved simplified eddy viscosity based $k-\epsilon-a$ models for variable density flows \cite{schwarzkopf2016}, better algebraic closures and more accurate sub-grid scale models.

The fruition of this potential of Reynolds stress models depends on the quality of the closures for the individual turbulence processes in the Reynolds Stress Modeling approach. Along with progress in modeling, this requires accurate, varied and detailed data from experimental investigations. Experimental studies have a symbiotic relationship with turbulence modeling. Data from such experiments can guide the development and testing of models. For example the experiments of \cite{lumley1977} pointed to a non-linear return to isotropy phenomenon in decaying turbulence. This led to the formulation of advanced slow pressure strain correlation models like \cite{ssmodel}. The shortcomings in models also guide the organization of new experiments. For example the drawbacks of turbulence models in rotation dominated mean flows led to the investigations of \cite{bns,bardina1985}. While established models are available for the evolution of turbulence processes there remain many questions about the model expressions and the closure coefficients. For example the closure coefficient values used in the rate of dissipation evolution model are varied between different studies in literature. Most of these studies use closure coefficient values that are well outside the range established by theoretical guidelines and experimental investigations. Similarly the form of the model expression used in pressure strain correlation models is also not universally accepted. The model of \cite{rotta1951} is linear in the Reynolds stress anisotropies, but the model of \cite{ssmodel} is non-linear with coefficients that are functions of the Reynolds stress invariants. While the models of \cite{rotta1951} and \cite{ssmodel} use a modeling basis consisting of the Reynolds stress anisotropies, the model of \cite{panda2017} uses additional tensors in the modeling expression. Using the experimental data from this study we evaluate these variabilities and make recommendations for improvement.  

In this investigation we study the canonical cases of Homogeneous Isotropic Turbulence (HIT) and Homogeneous Anisotropic Turbulence (HAT). HIT conditions are well replicated in experimental grid generated turbulence. Such tests are conducted in wind tunnels or water tanks where the grid is placed at the beginning of the test section. The rods in the grid interact with the flow through them leading to wakes. Just downstream of the grid, the wakes from individual rods interact with each other producing turbulence. If there is no externally imposed forcing downstream of the grid this turbulent kinetic energy is viscously dissipated at small scales leading to a decay in the velocity fluctuations. This turbulent velocity field becomes statistically isotropic at a distance of the order of 10-20 mesh lengths from the grid \cite{lumley1977}. Beyond this length this turbulent flow is statistically stationary with variation along the stream wise direction as the turbulence decays. The rate of energy decay is approximately equal to the viscous dissipation rate. Many authors have explored grid generated turbulence \cite{warhaft1978,le1985,comte1966,choi2001}. In addition to the insight into the decay of turbulence such studies provide data for benchmarking and calibrating turbulence models.

HAT conditions are imposed by using passage of the turbulent flow through an area change in the flow duct. Axisymmetric contraction increases the turbulent velocity fluctuations along the transverse directions. \cite{choi2001} have studied wind-tunnel turbulence experimentally and explored plane distortion, axisymmetric expansion and contraction to introduce anisotropy in grid turbulence. \cite{murzyn2005} investigated the grid generated turbulence experimentally using a water tank and have studied the evolution of turbulence kinetic energy, dissipation rate and other flow parameters. \cite{ayyalasomayajula2006} experimentally investigated grid-generated turbulence subjected to axisymmetric strain and indicated that single-point turbulence models may not be adequate to describe the relaxation of the turbulence towards an isotropic state. In a very important investigation \cite{torrano2015} studied the properties of turbulence downstream of a passive grid. This experimental data was used to evaluate the accuracy of eddy viscosity models and suggest optimal values of model coefficients. Such studies provide essential guidance for the limitations and development of improved turbulence models. 

In spite of these investigations few researchers have investigated the detailed evolution of Reynolds stress anisotropies near the grid at a large range of grid Reynolds numbers. This represents one of the novel contributions of this paper. This paper presents both experimental and numerical analysis of grid generated turbulence with and without mean strain. In the experimental portion of this paper, we investigate the degree of the anisotropy of the time averaged turbulent field at different grid Reynolds numbers with and without mean strain. The evolution of pressure coefficient, turbulent kinetic energy and Reynolds stresses are plotted downstream of the grid for having clear picture of turbulence structure near the grid. In the numerical analysis, we use this experimental data to analyze the closure coefficients of the rate of dissipation model and the pressure strain correlation models used in Reynolds Stress Modeling simulations. Based on this analysis we make recommendations for optimal values of the coefficients for studies of grid-generated turbulence.

\section{Experimental and modeling details}
\begin{figure}
\begin{centering}
\includegraphics[width=0.7\textwidth]{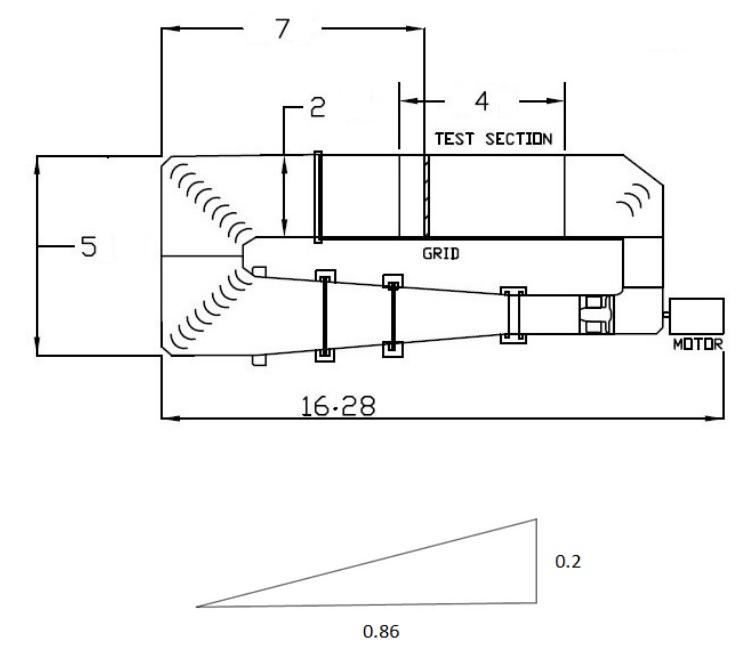}
\caption{Schematic of the recirculating water tank and the wedge, all dimensions are in meter, the width of the wedge is same as width of the water tank  \label{fig:1}}
\end{centering}
\end{figure}

The experiments for this paper were conducted in the recirculating water tank at the department of Ocean Engineering and Naval Architecture, IIT Kharagpur. Side walls of the water tank are made up of glass. The schematic of the experimental apparatus is shown in figure \ref{fig:1}. The water is recirculated by a pump, the rpm of the pump is controlled by an electrical control unit. A mean flow velocity of $1 m/s$ is achievable for a water depth of $0.8$ meter. The water tank has width 2 meters and depth 1.5 meter. The grids were placed immediately preceding the test section through a grid holder. The depth of water was 0.8 meter for all the cases of the experiments. Turbulence was generated by using a grid made up of cylindrical pipes. The diameter of the pipes used was 0.025 meter. The mesh length of the grids (M) was 10cm. The rigidity of the grid was calculated as 0.43 by using equation (1) as described in \cite{comte1966}
\begin{equation}
\sigma={d_b/M}{(2-d_b/M)}
\end{equation}

Reynolds number based on the grid mesh size \cite{nagata2008} is calculated as:
\begin{equation}
Re_M={\overline{U}M}{/\nu},
\end{equation}
here $M$ is the mesh size, $\overline{U}$ is the inflow velocity and $\nu$ is the kinematic viscosity of water. Experiments were conducted at three different grid Reynolds numbers, $Re_M$= 25000, 32000 and 39000.

Wedges were used downstream of the mesh for contracting the flow. These were fixed at a distance of 0.6 meter downstream of the grid. The detailed dimensions of the wedges is shown in figure 1. x is the main flow direction (at the grid position, x=0), y is the transverse direction and z is the vertical direction. $U, V$ and $W$ are the horizontal, transverse and vertical velocity components.
An Acoustic Doppler Velocimeter is used in our experiment to measure instantaneous velocity components at different downstream locations of the grid. An ADV measures three-dimensional flow velocities using doppler shift principle. Main components of the instrument are a sound emitter, three sound receivers and a signal conditioning electronic module. A detailed overview of this technique can be found in \cite{garcia2007}.

\subsection{Data Analysis}
The data collected from the ADV were decomposed into Mean and fluctuating velocities as
\begin{equation}
v_i=\overline{v_i}+u_i
\end{equation}
$\overline{v}$ and $u$ can be calculated from the following formula
\begin{equation}
\overline{v}=1/n\sum_{i=1}^{n} U_i
\end{equation}

\begin{equation}
u=\sqrt{1/n\sum_{i=1}^{n} (v_i-\overline{v})^2}
\end{equation}
Since turbulence is considered as eddying motion of fluid , secondary stresses appear in the fluid and those stresses are known as Reynolds stresses, which is a second order tensor having nine components, out of which six are independent. Diagonal components are called as normal stresses and the off diagonal components are called as shear stresses. The turbulent kinetic energy is defined as $k=\frac{{\overline{u}}^2+{\overline{v}}^2+{\overline{w}}^2}{2}$.

\subsection{Modeling details}
The Reynolds stress transport equation has the form:
\begin{equation}
\begin{split}
\partial_{t} \overline{u_iu_j}+U_k \frac{\partial \overline{u_iu_j}}{\partial x_k}&=P_{ij}-\frac{\partial T_{ijk}}{\partial x_k}-\epsilon_{ij}+\phi_{ij},\\
\mbox{where},\\ 
P_{ij}=-\overline{u_ku_j}\frac{\partial U_i}{\partial x_k}-\overline{u_iu_k}\frac{\partial U_j}{\partial x_k}, & T_{kij}=\overline{u_iu_ju_k}-\nu \frac{\partial \overline{u_iu_j}}{\partial{x_k}}+\delta_{jk}\overline{ u_i \frac{p}{\rho}}+\delta_{ik}\overline{ u_j \frac{p}{\rho}}\\
,\epsilon_{ij}=-2\nu\overline{\frac{\partial u_i}{\partial x_k}\frac{\partial u_j}{\partial x_k}},& \phi_{ij}= \overline{\frac{p}{\rho}(\frac{\partial u_i}{\partial x_j}+\frac{\partial u_j}{\partial x_i})}\\
\end{split}
\end{equation}
$P_{ij}$ denotes the production of turbulence, $T_{ijk}$ is the diffusive transport, $\epsilon_{ij}$ is the dissipation rate tensor and $\phi_{ij}$ is the pressure strain correlation. The pressure fluctuations are governed by a Poisson equation:
\begin{equation}
\frac{1}{\rho}{\nabla}^2(p)=-2\frac{\partial{U}_j}{\partial{x}_i}\frac{\partial{u}_i}{\partial{x}_j}-\frac{\partial^2 u_iu_j}{\partial x_i \partial x_j}
\end{equation}

The fluctuating pressure term is split into a slow and rapid pressure term $p=p^S+p^R$. Slow and rapid pressure fluctuations satisfy the following equations
\begin{equation}
\frac{1}{\rho}{\nabla}^2(p^S)=-\frac{\partial^2}{\partial x_i \partial x_j}{(u_iu_j-\overline {u_iu_j})}
\end{equation}
\begin{equation}
\frac{1}{\rho}{\nabla}^2(p^R)=-2\frac{\partial{U}_j}{\partial{x}_i}\frac{\partial{u}_i}{\partial{x}_j}
\end{equation}
It can be seen that the slow pressure term accounts for the non-linear interactions in the fluctuating velocity field and the rapid pressure term accounts for the linear interactions. A general solution for $\phi_{ij}$ can be obtained by applying Green's theorem to equation (7):
\begin{equation}
\phi_{ij}=\frac{1}{4\pi}\int_{-\infty}^{\infty} \overline{\frac{\partial{u}^*_k}{\partial{x_l^*}}\frac{\partial{u}^*_l}{\partial{x_k^*}}(\frac{\partial{u}_i}{\partial{x_j}}+\frac{\partial{u}_j}{\partial{x_i}})}
+2G_{kl}\overline{\frac{\partial{u}^*_l}{\partial{x_k^*}}(\frac{\partial{u}_i}{\partial{x_j}}+\frac{\partial{u}_j}{\partial{x_i}})}\frac{dVol^*}{|{x_n-x_n^*}|}
\end{equation}
The volume element of the corresponding integration is $dVol^*$. Instead of an analytical approach, the pressure strain correlation is modeled using rational mechanics approach. The rapid term can be modeled by assuming the length scale of mean velocity gradient is much larger than the turbulent length scale and is written in terms of a fourth rank tensor \cite{pope2000}
\begin{equation}
\phi_{ij}^R=4k\frac{\partial{U}_l}{\partial{x_k}}(M_{kjil}+M_{ikjl})
\end{equation}
where, 
\begin{equation}
M_{ijpq}=\frac{-1}{8\pi k}\int \frac{1}{r} \frac {\partial^2 R_{ij}(r)}{\partial r_p \partial r_p}dr
\end{equation}
where, $R_{ij}(r)=\langle u_i(x)u_j(x+r) \rangle$

For homogeneous turbulence the complete pressure strain correlation can be written as
\begin{equation}
\phi_{ij}=\epsilon A_{ij}(b)+kM_{ijkl}(b)\frac{\partial\overline {v}_k}{\partial{x_l}}
\end{equation}
The most general form of slow pressure strain correlation is given by
\begin{equation}
\phi^{S}_{ij}=\beta_1 b_{ij} + \beta_2 (b_{ik}b_{kj}- \frac{1}{3}II_b \delta_{ij})
\end{equation}
Established slow pressure strain correlation models including the models of \cite{rotta1951} and \cite{ssmodel} use this general expression. Considering the rapid pressure strain correlation, the linear form of the model expression is
\begin{equation}
\frac{\phi^{R}_{ij}}{k}=C_2 S_{ij} +C_3 (b_{ik}S_{jk}+b_{jk}S_{ik}-\frac{2}{3}b_{mn}S_{mn}\delta_{ij})+  
C_4 (b_{ik}W_{jk} + b_{jk}W_{ik})
\end{equation}
Here $b_{ij}=\frac{\overline{u_iu_j}}{2k}-\frac{\delta_{ij}}{3}$ is the Reynolds stress anisotropy tensor, $S_{ij}$ is the mean rate of strain and $W_{ij}$ is the mean rate of rotation. Rapid pressure strain correlation models like the models of \cite{lrr} and \cite{mishra6} use this general expression. 




\section{Experimental Results}

\begin{figure}
\includegraphics[width=0.5\textwidth]{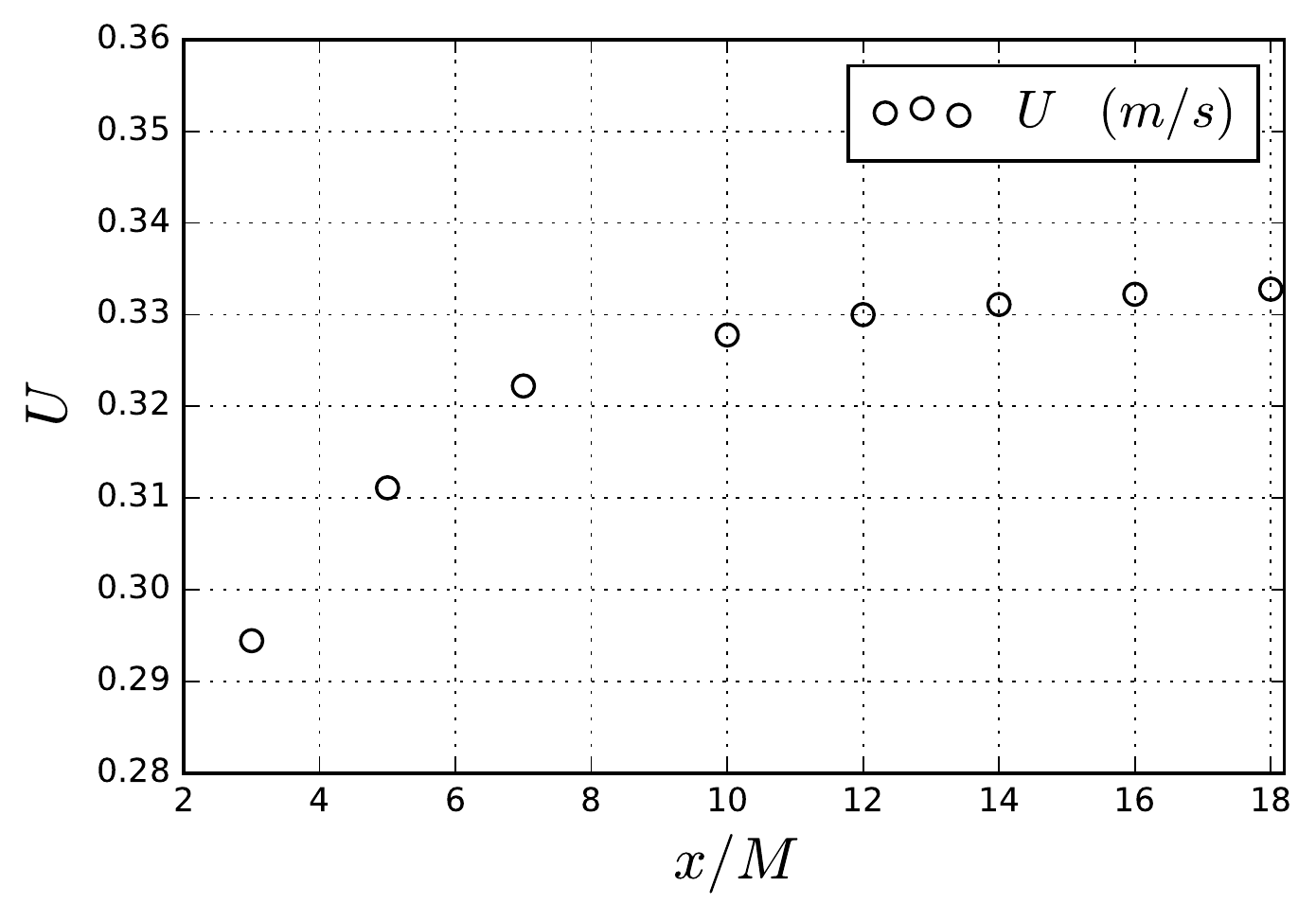}
\includegraphics[width=0.5\textwidth]{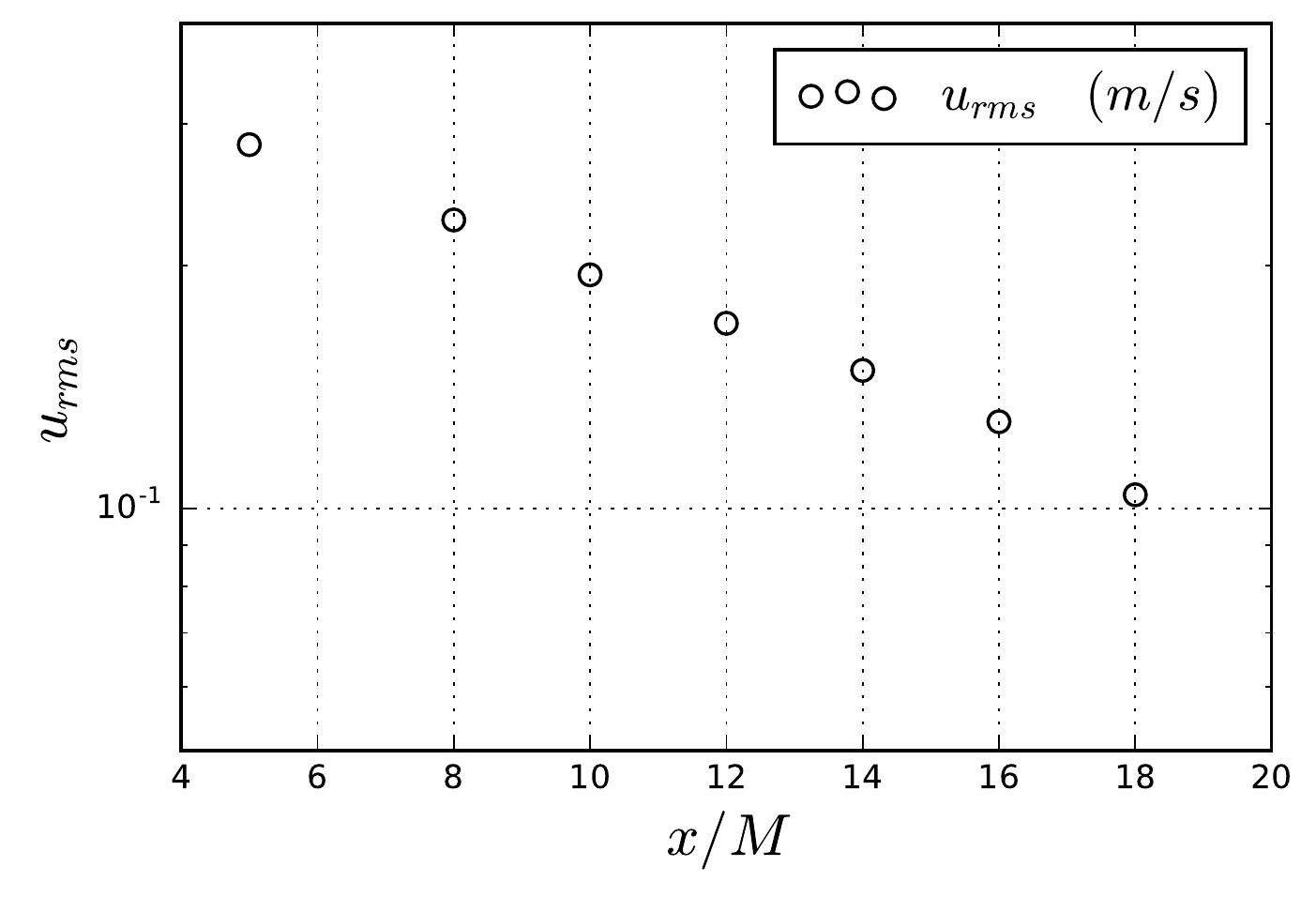}
\caption{Evolution of the streamwise mean velocity $U$ and the fluctuating velocity $u_{rms}$ downstream of the mesh. \label{fig:2}}
\end{figure}

In the experimental results we report the Reynolds stress anisotropies and the decay of the turbulence kinetic energy downstream of the mesh for a range of different Reynolds numbers. In figure \ref{fig:2} we show the evolution of the streamwise mean velocity $U$ and the fluctuating velocity $u_{rms}$ downstream of the mesh for $Re_M =25000$. There is a gradual increase in the streamwise mean velocity due to the development of the boundary layers along the configuration walls till it reaches its maximum value. The power law decay of the fluctuating velocity is shown in the second sub-figure. The exponent of decay for this case was calculated from the experimental data to be $1.35$. 

\begin{figure}
\begin{centering}
\includegraphics[width=0.6\textwidth]{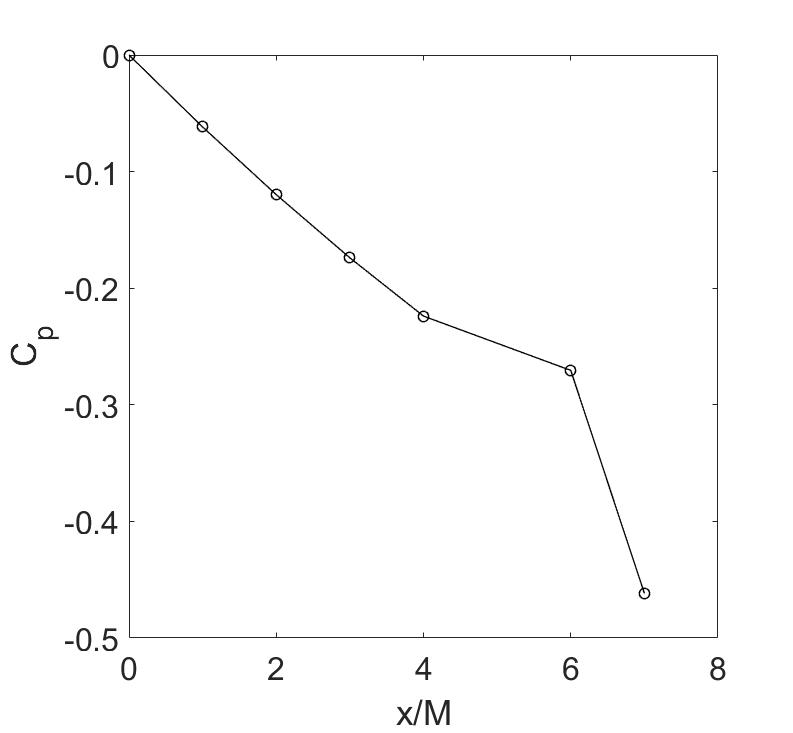}
\caption{Distribution of pressure coefficient along the streamwise direction for $Re_M=25000$, the reference pressure was taken at the beginning of the contraction, the downstream distance is non-dimensionalized by the wedge step height\label{fig:3}}
\end{centering}
\end{figure}

The pressure distribution is expressed in terms of the pressure coefficient $C_p$. For an incompressible flow $C_p=\frac {P-P_{ref}}{0.5\rho U_{ref}^2}$, where $p$ is the local static pressure, $p_{ref}$  
is the static pressure at the beginning of the contraction, the local free stream velocity is $U_{infinity}$ and $U_{ref}$ is the reference free stream velocity at the beginning of the contraction. In figure \ref{fig:3}, the evolution of the pressure coefficient along the stream wise direction is presented, since the contraction used in the experiment produces a favorable pressure gradient, a sharp decrease in pressure coefficient is observed along the stream wise direction. The downstream distance is normalized with respect to the step height of the contraction and $x/M =0$ represents the beginning of the contraction.

\begin{figure}
\center
\includegraphics[width=0.7\textwidth]{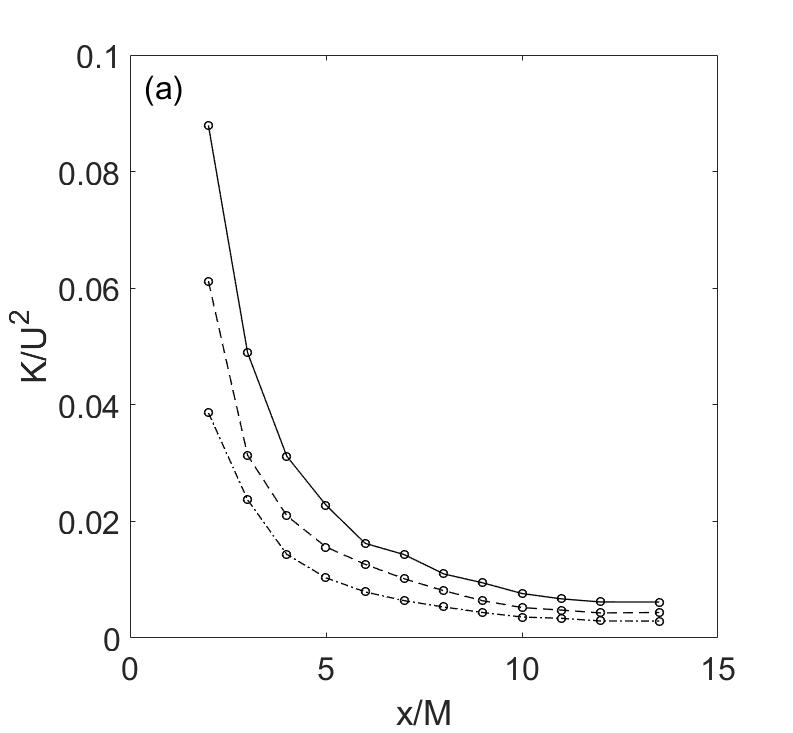}
\includegraphics[width=0.7\textwidth]{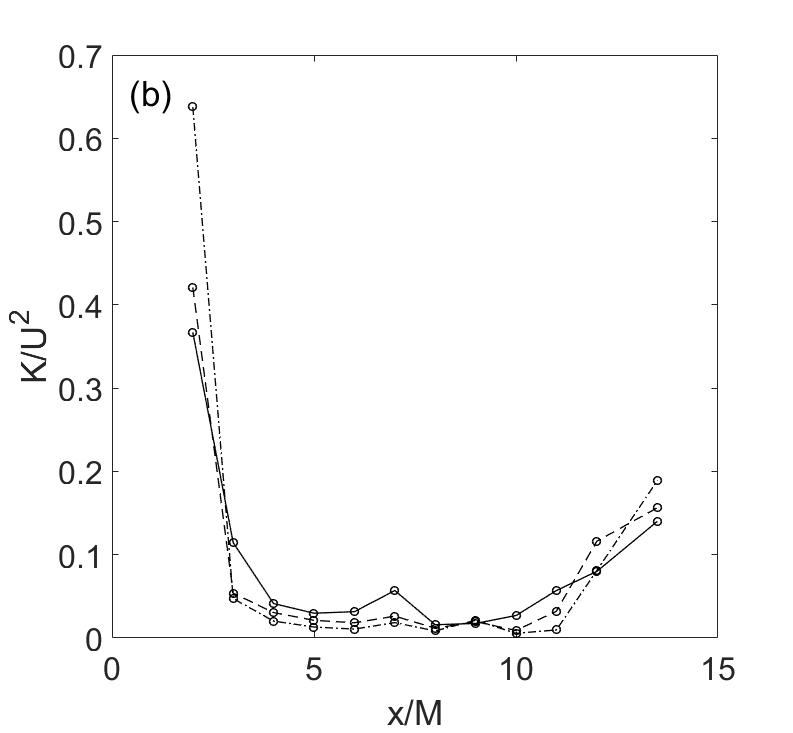}
\caption{Downstream evolution of turbulence kinetic energy, a) without mean strain b) with mean strain, solid lines corresponds to $Re_M$ of 39000, dashed lines 32000 and dashed dot lines 25000. \label{fig:4}}
\end{figure}

In figure \ref{fig:4} the turbulence kinetic energy evolution for three different grid Reynolds number is presented. It is observed in the absence of mean strain there is a sharp decay of turbulence kinetic energy and with increase in grid Reynolds number turbulence kinetic energy increases. However in presence of mean strain, an increase in turbulence kinetic energy just after the contraction is observed. Because of the favorable pressure gradient the turbulence kinetic energy near the grid increased five times on an average at all three grid Reynolds numbers.

\begin{figure}
\begin{centering}
\includegraphics[width=0.7\textwidth]{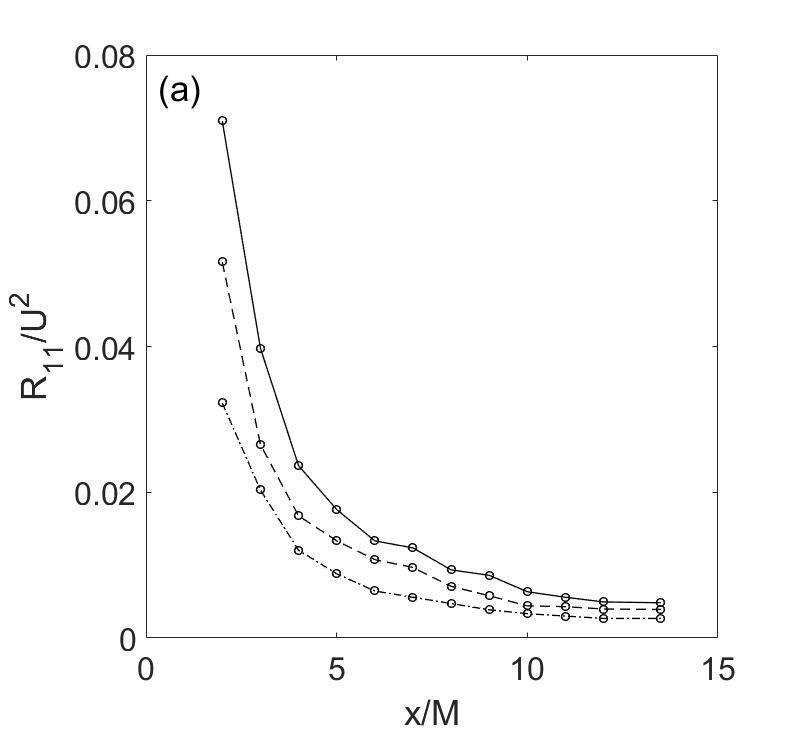}
\includegraphics[width=0.7\textwidth]{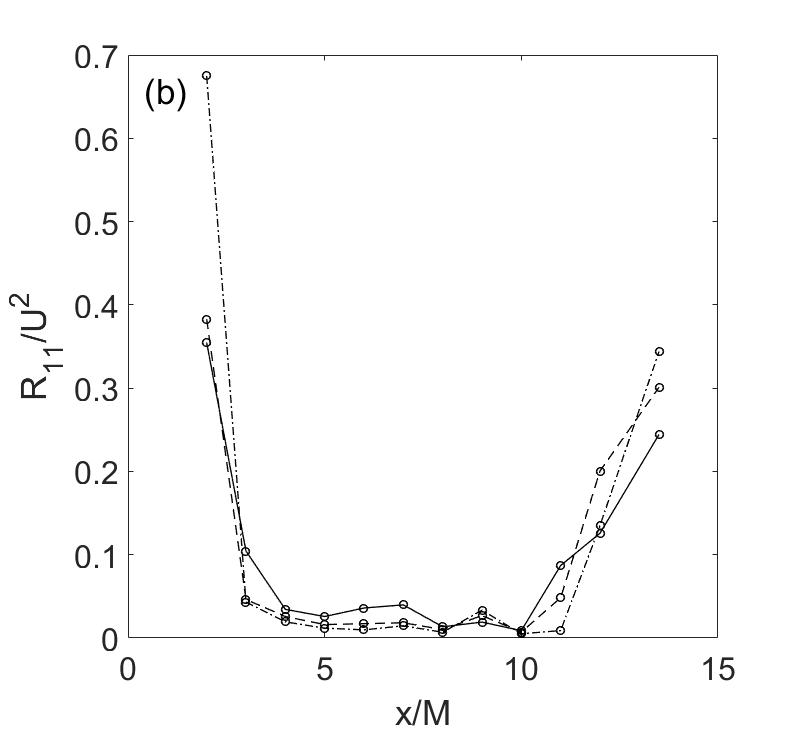}
\caption{Downstream evolution of Reynolds stress, a) without mean strain b) with mean strain, solid lines corresponds to $Re_M$ of 39000, dashed lines 32000 and dashed dot lines 25000. \label{fig:5}}
\end{centering}
\end{figure}

\begin{figure}
\begin{centering}
\includegraphics[width=0.7\textwidth]{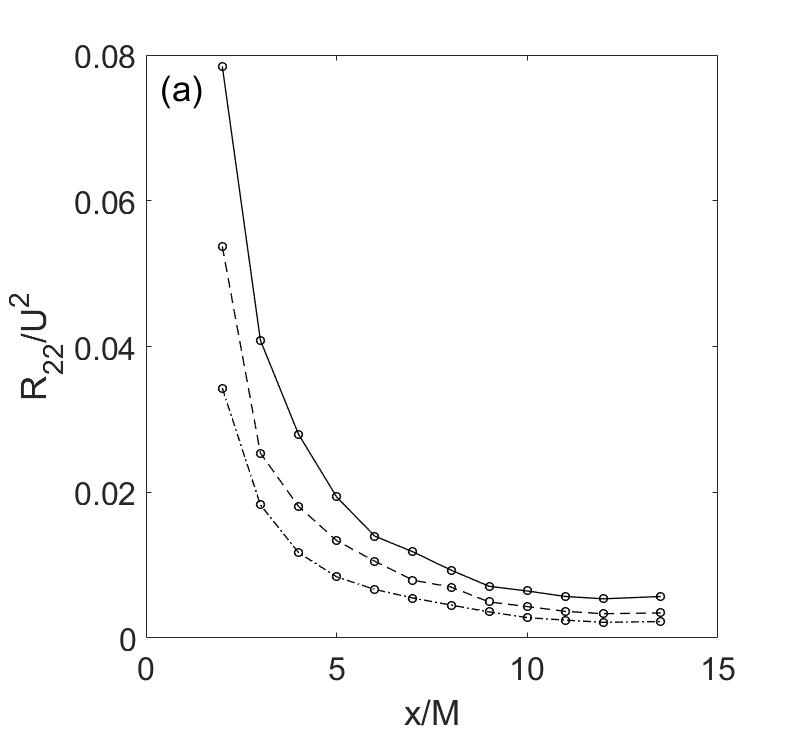}
\includegraphics[width=0.7\textwidth]{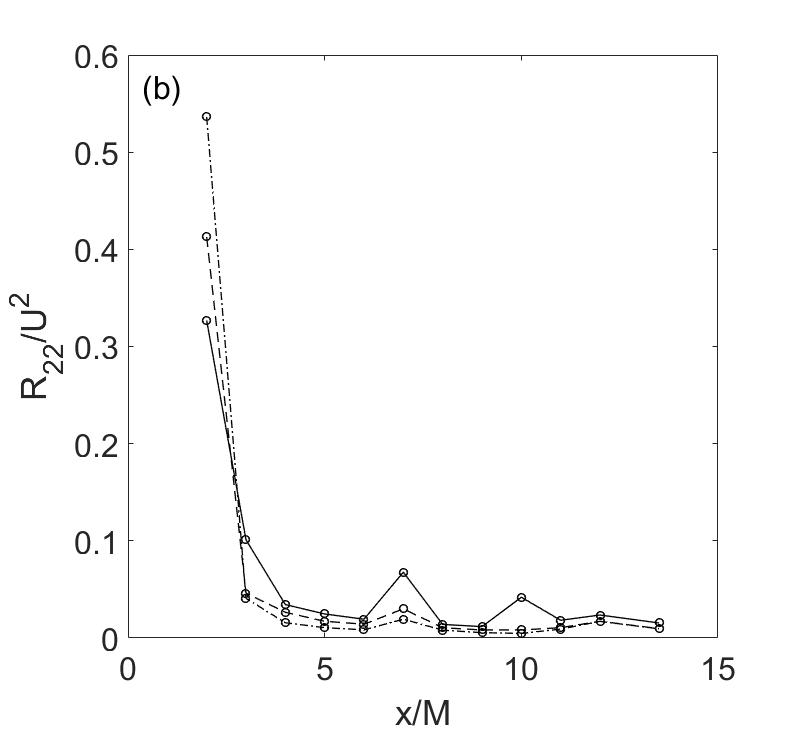}
\caption{Downstream evolution of Reynolds stress, a) without mean strain b) with mean strain, solid lines corresponds to $Re_M$ of 39000, dashed lines 32000 and dashed dot lines 25000. \label{fig:6}}
\end{centering}
\end{figure}

In figure \ref{fig:5} and \ref{fig:6}, the evolution of normal components of Reynolds stresses are shown. It is observed that the imposed strain has no effect on the distribution of Reynolds stresses along the transverse direction, but in longitudinal direction, there is an  increase in Reynolds stress components towards the end of the contraction.

\begin{figure}
\begin{centering}
\includegraphics[width=0.7\textwidth]{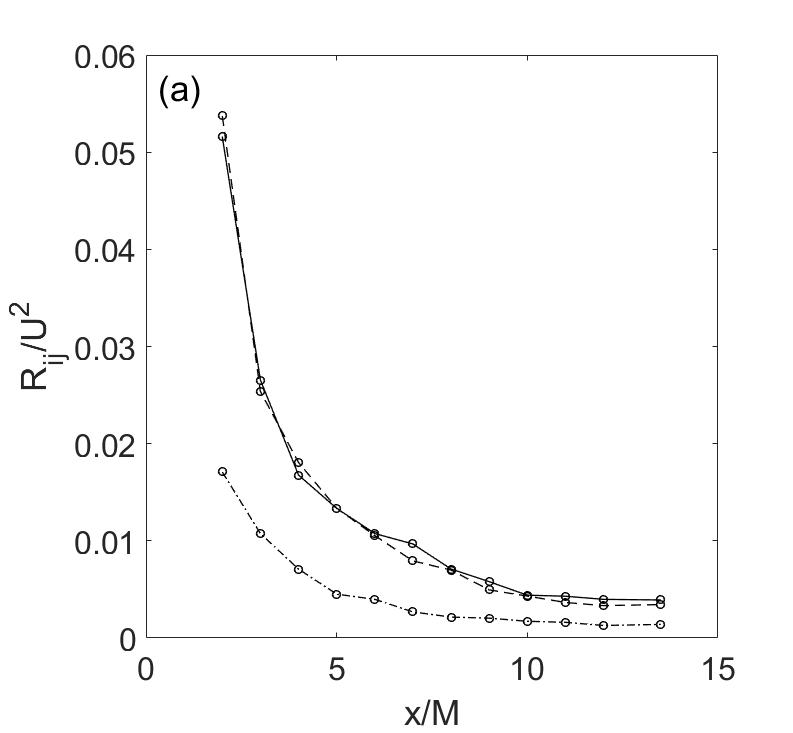}
\includegraphics[width=0.7\textwidth]{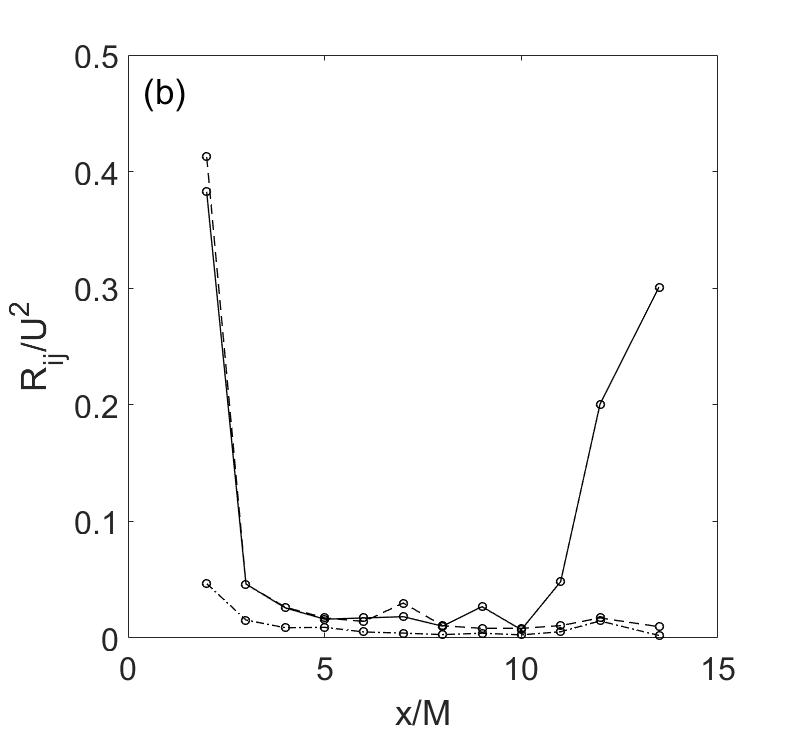}
\caption{Effect of mean strain on free stream turbulence, a) represents evolution of free stream turbulence without mean strain b) with mean strain. solid line represents $R_{11}$ ,dashed lines $R_{22}$ and dashed dot lines $R_{33}$ corresponding to $Re_M$ of 32000. \label{fig:7}}
\end{centering}
\end{figure}

The comparison of the evolution of Reynolds stresses at $Re_M=32,000$ is shown in figure \ref{fig:7}. It is shown in the figure that the imposed strain enhances the magnitude of Reynolds stress only in the longitudinal direction. 



\begin{figure}
\begin{centering}
\includegraphics[width=0.8\textwidth]{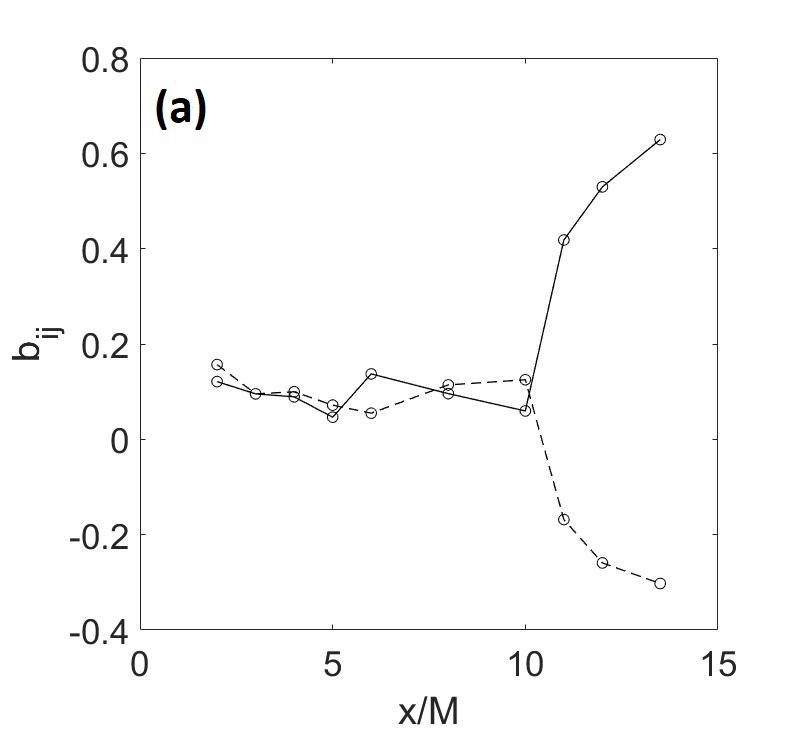}
\includegraphics[width=0.8\textwidth]{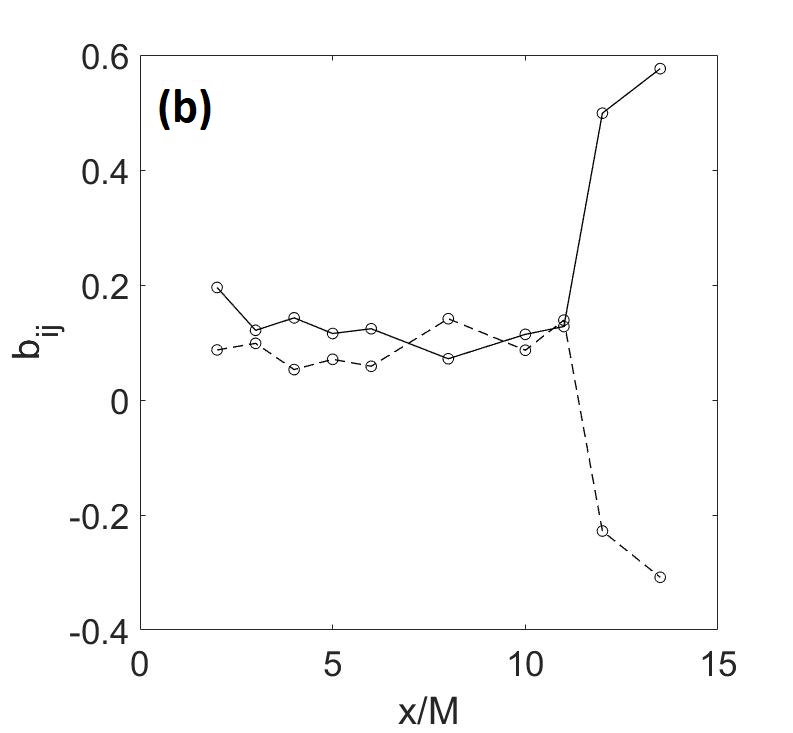}
\caption{$b_{ij}$ evolution under mean strain at $Re_M=$a)32000 b)39000. Solid lines show the longitudinal component, dashed lines the transverse normal component\label{fig:9}}
\end{centering}
\end{figure}

The downstream evolution of Reynolds stress anisotropies for grid with contraction case is shown in figure \ref{fig:9}. The imposed contraction leads to increase in the anisotropy of the turbulent flow field.  

\section{Numerical results and analysis}
\subsection{Analysis of the rate of dissipation model}
Experimental investigations of the decay of grid generated turbulence are essentially important, but there is a marked trend to investigate such flow cases using numerical simulations. Numerical simulations can provide idealized conditions for the experiment (that may only be met approximately in a real experiment). Numerical simulations also provide a large amount of detailed data which is free of measurement errors. Most such numerical simulations into the decay of turbulence use the Reynolds stress modeling approach.  

An important shortcoming for the Reynolds Stress Modeling approach is the approximate nature of the rate of dissipation equation. While the model equations for the evolution of the Reynolds stress anisotropy components are exact and based on the Reynolds stress transport equation the evolution equation for the rate of dissipation is empirically derived \cite{pope2000}. This model expression is 
\begin{equation}
\frac{D\epsilon}{Dt}=\frac{\partial }{\partial x_i} (\nu +\frac{\nu_t}{\sigma_k})\frac{\partial \epsilon}{\partial x_i} +C_{\epsilon 1} \frac{P\epsilon}{k} - C_{\epsilon 2} \frac{\epsilon^2}{k}
\end{equation}
The first term on the right hand side represents the diffusive transport of $\epsilon$. The second and third terms on the right side represent the generation of $\epsilon$ due
to vortex stretching and the destruction of $\epsilon$ by viscous action. The standard values for the closure coefficients are given by $\sigma_{\epsilon}=1.3$, $C_{\epsilon 1}=1.44$ and $C_{\epsilon 1}=1.92$, based on the constants determined by \cite{launder1974}. The value of the $C_{\epsilon 2}$ coefficient is calibrated to be in agreement with the power law decay observed in decaying turbulence. Here the decay exponent corresponds to the power law decay observed as $k(t)=k(t_0)(t/t_0)^{-n}$ and $\epsilon(t)=\epsilon(t_0)(t/t_0)^{-n-1}$. In terms of the decay exponent $n$ this is given by
\begin{equation}
n=\frac{1}{C_{\epsilon 2} -1}
\end{equation}

\begin{figure}
\begin{centering}
\includegraphics[width=0.7\textwidth]{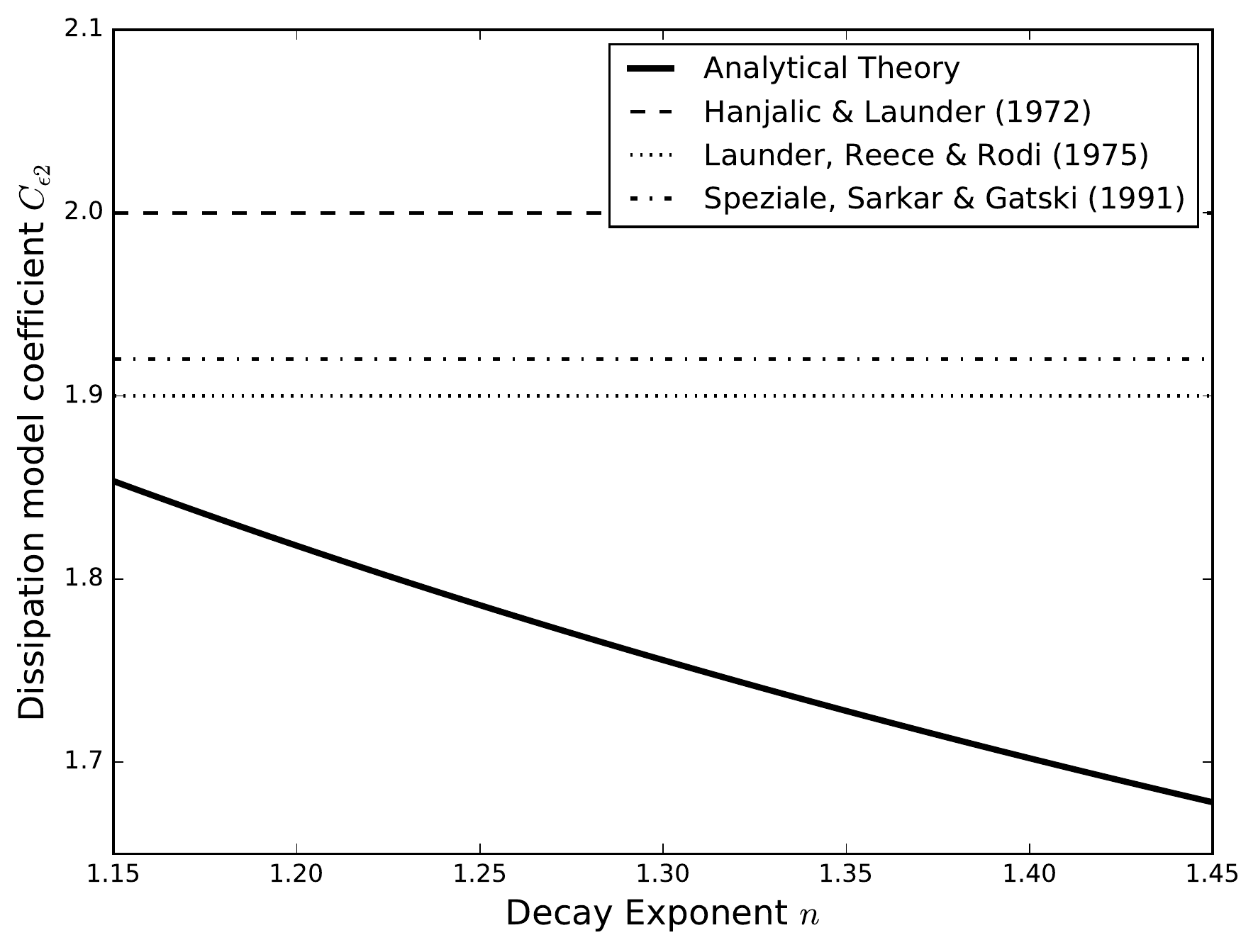}
\caption{Contrasting the relationship between $C_{\epsilon 2}$ and $n$ based on experimental studies (solid black line) and the values used in Reynolds Stress Modeling investigations. \label{fig:ce2}}
\end{centering}
\end{figure}

Most experimental investigations have found the decay exponent to lie in the range of $1.15-1.45$. This obligates the value of $C_{\epsilon 2}$ to approximately lie in the range $1.69-1.87$. However the values used in different models often lies well outside this bound. Based on \cite{batchelor1948}, \cite{hanjalic1972} chose $C_{\epsilon 2}=2.0$ to make the turbulent kinetic energy vary inversely with distance from the origin. Both the investigations of \cite{lrr} and \cite{ssmodel} changed it to $C_{\epsilon 2}=1.9$ so as to get faster rate of decay for their model simulations. \cite{ssg} chose to adopt the value of $C_{\epsilon 2}=1.92$ for better calibration of their model. Since then, different modeling investigations have used different values for the coefficient varying from $1.90$ to $2.0$. All these chosen values lie outside the range prescribed by experimental investigations and are often varying from one numerical investigation to another. In this context the value of the $C_{\epsilon 2}$ is important to ensure correct simulations and its determination may represent a hurdle for the Reynolds stress modeling approach. 

The values of the $C_{\epsilon 1}$ is chosen to match the steady state parameters in homogeneous turbulent shear flow. The form is given by $\frac{P}{\epsilon}=\frac{C_{\epsilon 2} -1}{C_{\epsilon 1} -1}$. It can be seen form this relationship that the choice of the value of the coefficient $C_{\epsilon 2}$ also in turn affects the value of the $C_{\epsilon 1}$ coefficient. Any errors in the values of $C_{\epsilon 2}$ will have a cascading effect and will affect the accuracy of the entire model. 

\begin{figure}
\begin{centering}
\includegraphics[width=0.7\textwidth]{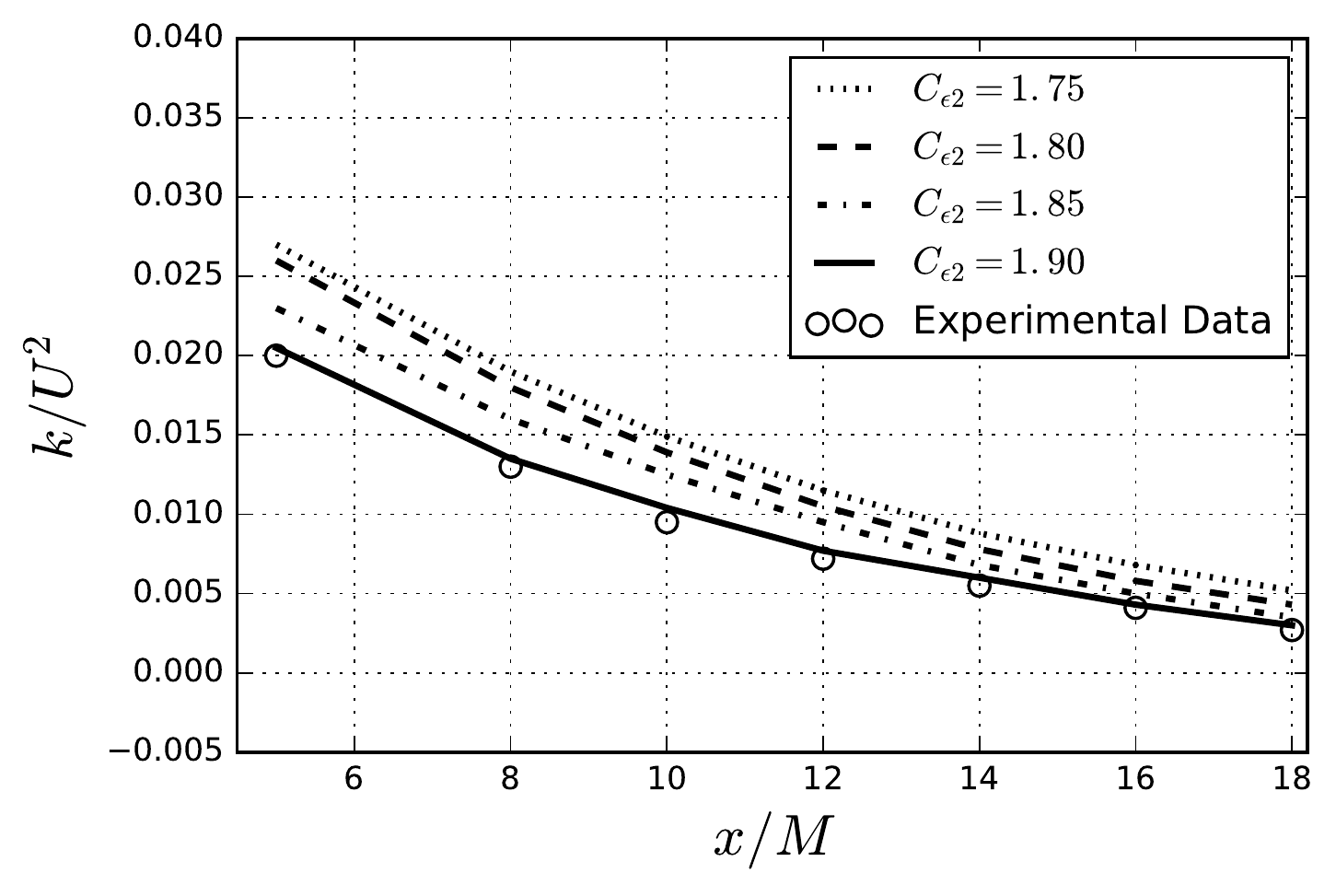}
\caption{Turbulent kinetic energy downstream of the mesh and the decay predicted by the model \cite{ssg} with different values of the coefficients for the rate of dissipation equation.\label{fig:ssg}}
\end{centering}
\end{figure}

In this section, we vary the value of $C_{\epsilon 2}$ while using different established Reynolds Stress Models to find the optimal value for this coefficient. The values of $C_{\epsilon 1}$ and $\sigma_{\epsilon}$ ($\sigma_{\epsilon}=\frac{\kappa^2}{\sqrt{C_\mu}(C_{\epsilon 2}-C_{\epsilon 1}}$ where the Von Karman constant is $\kappa=0.41$) are determined by their relationship with $C_{\epsilon 2}$. While we have done this investigation for a large number of $C_{\epsilon 2}$ values, we show the results for four values $C_{\epsilon 2}=1.75, 1.80, 1.85, 1.90$. Only the first three values are in the range allowed by experimental data. Mesh independence studies were carried out for this case and are reported in \cite{panda2017}.

In figure \ref{fig:ssg} we show the the decay of turbulent kinetic energy downstream of the mesh predicted by the model of \cite{ssg}. The rate of decay of turbulent kinetic energy is captured well by the model predictions. At $C_{\epsilon 2}=1.75$ (corresponding to the decay exponent calculated from experimental data) the model over predicts the value of the turbulent kinetic energy. Increasing the value of $C_{\epsilon 2}$ leads to improvement in the prediction of the turbulent kinetic energy downstream of the mesh. At $C_{\epsilon 2}=1.90$ we get the best agreement with the experimental data. This value of $C_{\epsilon 2}$ is not in agreement with the decay exponent calculated from experimental data and is outside the range  prescribed by experimental investigations in literature.

\begin{figure}
\begin{centering}
\includegraphics[width=0.7\textwidth]{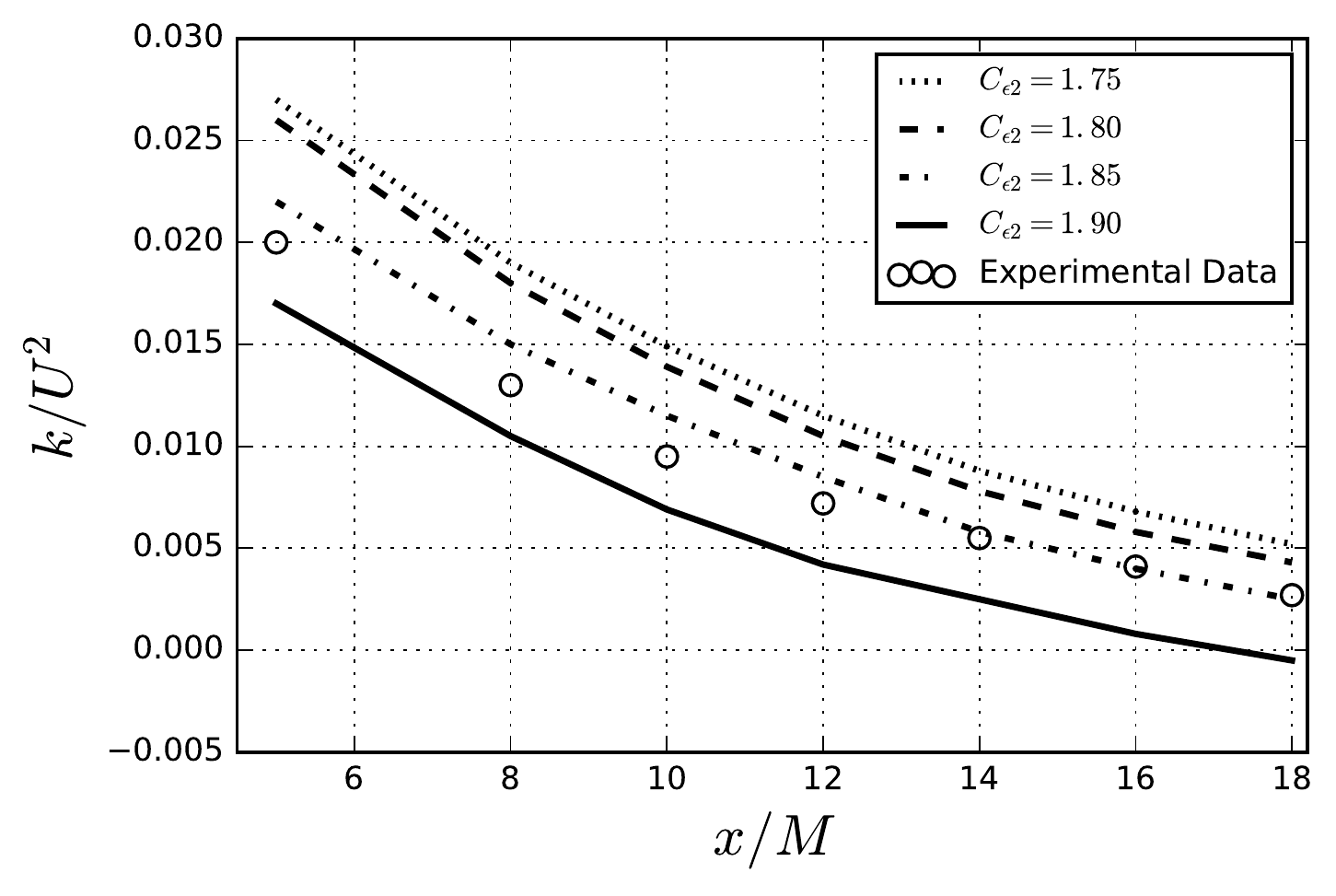}
\caption{Turbulent kinetic energy downstream of the mesh and the decay predicted by the model of \cite{lrr} with different values of the coefficients for the rate of dissipation equation.\label{fig:lrr}}
\end{centering}
\end{figure}

In figure \ref{fig:lrr} we show the the decay of turbulent kinetic energy downstream of the mesh predicted by the model of \cite{lrr}. In this case we see that at $C_{\epsilon 2}=1.90$ the dissipation downstream of the mesh is over-predicted. At $C_{\epsilon 2}=1.85$ the agreement between the experimental data and the model predictions is much better. 

Such variability in the coefficient values for turbulence models arises due to the empirical nature of the modeling expression and the modeling methodology where the coefficient values are calibrated using limited data from select experiments. To highlight such variability \cite{cheung2011} have developed different models calibrated against experimental data that show significant variation in the values of the coefficients. \cite{edeling2014} and \cite{edeling2014predictive} have studied the parameter variability across flows for the $k-\epsilon$ model. With respect to Reynolds stress modeling \cite{mishra2015epistemic} and \cite{mishra2016sensitivity} have shown that in homogeneous turbulence there can be significant variation in turbulence evolution if the Reynolds stresses are assumed to completely describe the state of the turbulent flow field.

This disagreement between theoretical analysis with experimental data and the numerical results against experimental data observed in this paper may be arising due to the empirical nature of the rate of dissipation evolution equation. In this case we did not vary the coefficient values of the model of \cite{ssg} and \cite{lrr}. These model coefficients are calibrated for homogeneous shear flows. Variation in these values may be advantageous because it would sample from all the degrees of freedom in the system of equations. 

From this analysis it is clear that the optimal values of the closure coefficients of the rate of dissipation equation depend on the flow being simulated and also on the pressure strain correlation model used. This represents a hurdle in numerical simulations as the rate of dissipation equation should be ideally independent of other models. As a compromise, we recommend the value of $C_{\epsilon 2}=1.87$. This determines the values of $C_{\epsilon 1}=1.42$ and $\sigma_{\epsilon}=1.25$. For these values both the models of \cite{ssg} or \cite{lrr} give acceptable agreement with experimental data. This value is also contained inside the range recommended by most experimental investigations in literature.

\subsection{Analysis of the Pressure strain correlation model}
In our analysis of the pressure strain correlation model we focus on the very popular LRR model of \cite{lrr}. The form of this model is given by
\begin{equation}
\begin{split}
\phi_{ij}=&-(C^{0}_1 \epsilon + C^{1}_{1}P)b_{ij}+C_2 kS_{ij} +C_3 k(b_{ik}S_{jk}+b_{jk}S_{ik}-\frac{2}{3}b_{mn}S_{mn}\delta_{ij})\\  
&+C_4 k(b_{ik}W_{jk} + b_{jk}W_{ik})
\end{split}
\end{equation}
The closure coefficients are given as $C^{0}_1=3$, $C^{1}_1=0$, $C_2=0.8$, $C_3=1.75$ and $C_4=1.31$. This model can be thought of as the summation of the slow pressure strain correlation model of \cite{rotta1951} with a rapid pressure strain correlation model developed by \cite{lrr}.

Inspite of its popularity and widespread use there are many questions raised in literature regarding the values of its coefficients. The closure coefficient values of \cite{lrr} were calibrated using experimental data using simple turbulent flows. In \cite{lrr} the experiments used were restricted to the low shear experiment of \cite{champagne1970}. Different investigators like \cite{jones1984} used data from other turbulent flows to re-calibrate the coefficients of this model and determined different values of the closure coefficients. \cite{mishra1} have analyzed this form of the pressure strain correlation model expression and have recommended that the closure coefficients be explicit function of the mean rate of strain and mean rate of rotation tensors. This would make the values of the closure coefficients vary across different flows for the same model. 

In our analysis of the pressure strain correlation model of \cite{lrr} we analyze the model for the slow pressure strain correlation in isolation first. This is given by $\phi_{ij}=-C^{0}_1 \epsilon b_{ij}$ and is equal to the return to isotropy model of \cite{rotta1951}. For assessment of this pressure strain correlation models, the downstream distance relative to the start point of contraction was measured by the transit time of the turbulence advection from the beginning of the contraction to a given stream wise position , $x$ \cite{hearst2014}:
\begin{equation}
t = \int_{0}^{z}\frac{1}{U(x)}dx
\end{equation}
where $x$ is the dummy integration variable and $U(x)$ is the local mean velocity at a position $z$. The experimentally calculated value for the initial value of the $S^*=\frac{Sk_0}{\epsilon_0}=1.43$ is used for the simulations. 

\begin{figure}
\begin{centering}
\includegraphics[width=0.6\textwidth]{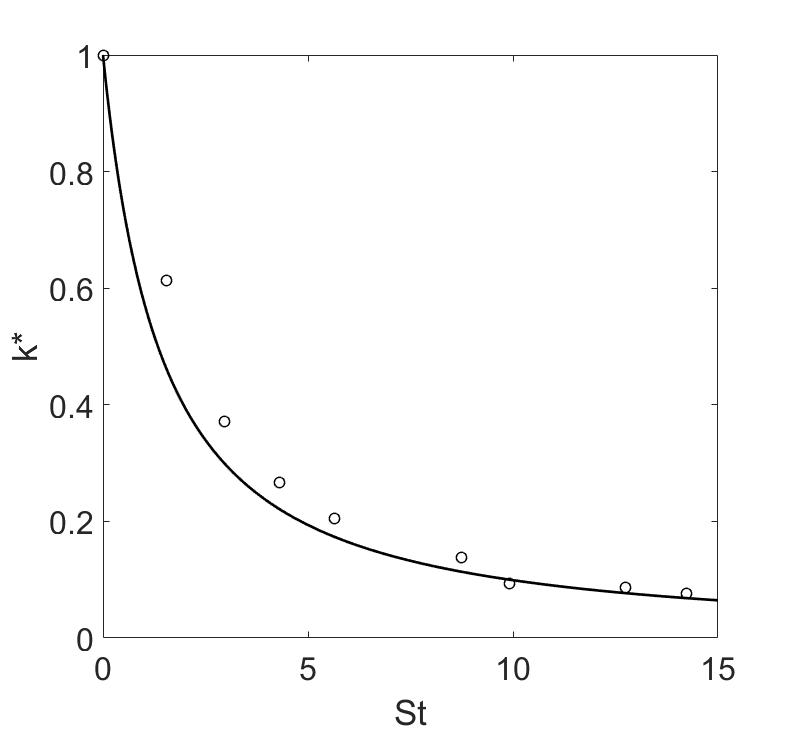}
\caption{Turbulence kinetic energy evolution for decaying grid generated turbulence, $S^*=1.43$. Predictions of \cite{rotta1951} are shown by solid line, results at $Re_M=25000$ by unfilled circles.\label{fig:lrr1}}
\end{centering}
\end{figure}

As can be seen in figure \ref{fig:lrr1} there is very good agreement between experimental data and model prediction when the linear interactions between the mean velocity field and the fluctuating velocity field are absent. This indicates that the slow pressure strain correlation model is adequate for simulation of grid generated decaying turbulence. 

Considering the rapid pressure strain correlation model given by the form
\begin{equation}
\phi^{R}_{ij}=C_2 kS_{ij} +C_3 k(b_{ik}S_{jk}+b_{jk}S_{ik}-\frac{2}{3}b_{mn}S_{mn}\delta_{ij})
+C_4 k(b_{ik}W_{jk} + b_{jk}W_{ik})
\end{equation}
There are 3 closure coefficients representing three potential degrees of freedom. However the value of the $C_2$ coefficient is fixed by the analytical Crow Constraint \cite{crow1968}. The other two coefficients are related to each other as $\frac{C_3}{6}+\frac{3C_4}{7}=\frac{8}{7}$ to maintain symmetry conditions on the $M_{ijkl}$ tensor \cite{pope2000}. Because of this there is just one degree of freedom in the coefficient values of the rapid pressure strain correlation model of \cite{lrr}. We choose to vary this degree of freedom to explore the optimal value for grid generated turbulence with and without the effects of mean straining. 

\begin{figure}
\begin{centering}
\includegraphics[width=0.7\textwidth]{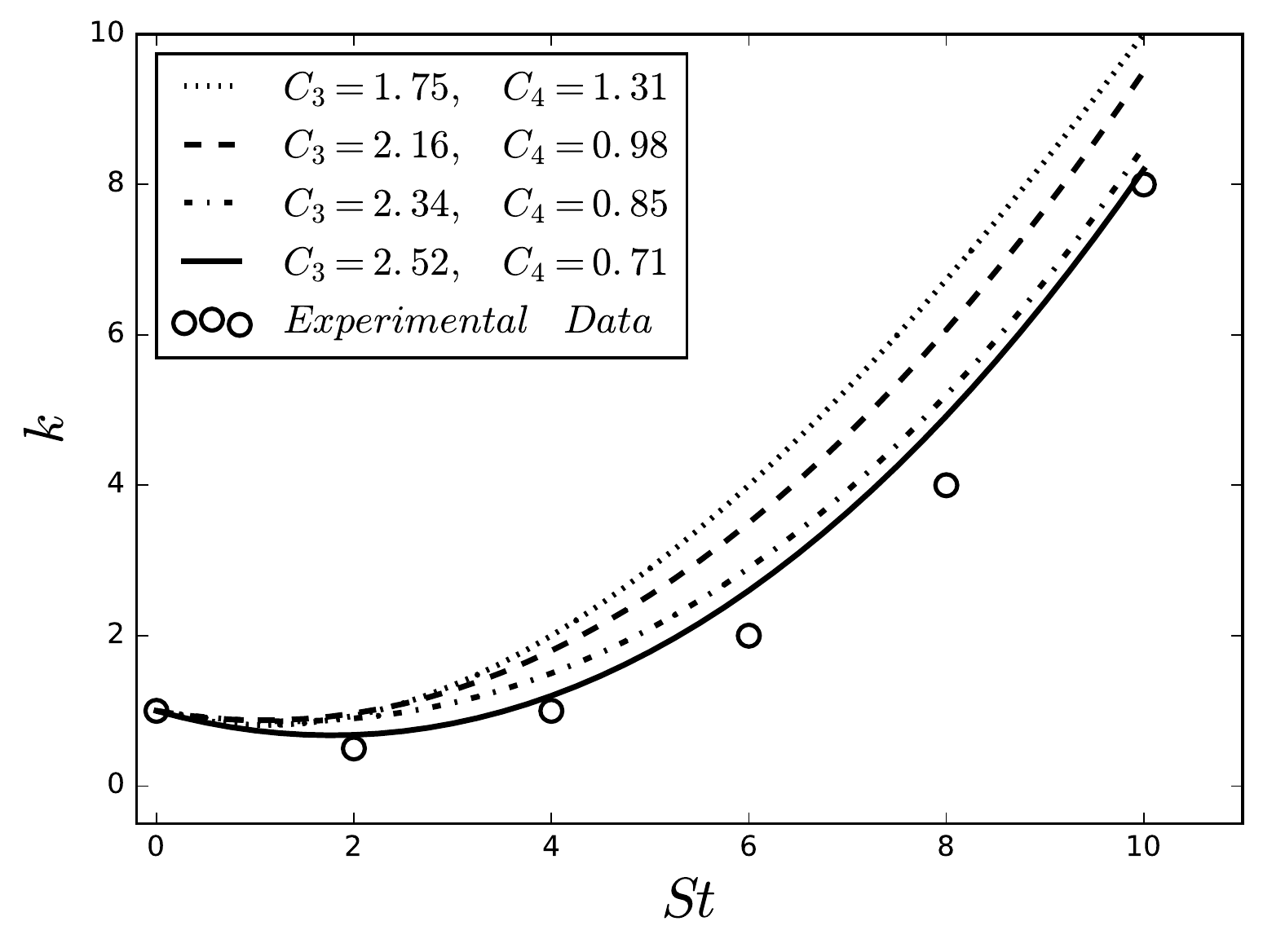}
\caption{Turbulence kinetic evolution for grid generated turbulence under axisymmetric contraction. Results at $Re_M=25000$ are shown as unfilled circles.\label{fig:lrr2}}
\end{centering}
\end{figure}

In figure \ref{fig:lrr2} we show the predictions of LRR model variants for the evolution of turbulent kinetic energy. The model coefficients can be expressed in terms of a closure coefficient of the $M_{ijkl}$ tensor, $\alpha$ as $C_3=-6\alpha$ and $C_4=\frac{2}{3}(4+7\alpha)$. We vary the value of $\alpha$ from $-0.29$ (corresponding to the model coefficient values in \cite{lrr}) to $-0.45$. As can be seen in the figure the decrease in value of $\alpha$ leads to improved predictions till $\alpha=-0.42$. Based on our analysis we recommend the value of $\alpha=-0.42$ (or $C_3=2.52,\quad  C_4=0.71$) for investigations grid generated turbulence undergoing mean staining effects.




\section{Conclusions}
In this paper we carried out experimental and numerical analysis of grid generated turbulence with and without mean strain. We conduct a series of experiments on decaying grid generated turbulence and grid turbulence with mean strain. Experimental data of turbulence statistics including Reynolds stress anisotropies is collected and analyzed. The decay of turbulent kinetic energy is mainly concentrated in the near grid region ($x/M <20$) in all three Reynolds numbers. The anisotropy of the flow field was presented for both the decaying grid turbulence and flow with mean strain. The experimental data was used to calibrate the coefficients of the rate of dissipation model and the pressure strain correlation models used in Reynolds Stress Modeling. For both models we recommend values of coefficients that should be used for experimental studies of grid generated turbulence.

In related future work we are using active grids to generate more varied data sets that cover a wider range of parameters including $\frac{Sk}{\epsilon}$, $Re_M$, $S$, $W$ and different initial conditions for the turbulent velocity field. This data will be used to generate probability distribution functions for the values of the closure coefficients that may be useful for Bayesian investigations into the variability of the values of these model coefficients.  

\bibliographystyle{elsarticle-num}
\bibliography{asme2e.bib}

\end{document}